\documentclass[review]{elsarticle}

\usepackage{lineno,hyperref}
\usepackage{subfigure}
\usepackage{longtable}
\usepackage{xcolor}
\modulolinenumbers[5]

\journal{Journal of \LaTeX\ Templates}

\begin{document}

\begin{frontmatter}

\title{The Anatomy of Brexit Debate on Facebook}

\author{Michela Del Vicario\fnref{myfootnote}}\author{Fabiana Zollo\fnref{myfootnote}}
\author{Guido Caldarelli}
\address{CSSLab, IMT School for Advanced Studies, Lucca, Italy}
\fntext[myfootnote]{These authors contributed equally to this work.}
\author{Antonio Scala}
\address{ICS CNR, Italy}
\author{Walter Quattrociocchi}
\address{CSSLab, IMT IMT School for Advanced Studies, Lucca, Italy}




\begin{abstract}
Nowadays users get informed and shape their opinion through social media. However, the disintermediated access to contents does not guarantee quality of information. Selective exposure and confirmation bias, indeed, have been shown to play a pivotal role in content consumption and information spreading. Users tend to select information adhering (and reinforcing) their worldview and to ignore dissenting information. This pattern elicits the formation of polarized groups -- i.e., echo chambers -- where the interaction with like-minded people might even reinforce polarization.
In this work we address news consumption around Brexit in UK on Facebook. In particular, we perform a massive analysis on more than 1 Million users interacting with Brexit related posts from the main news providers between January and July 2016. We show that consumption patterns elicit the emergence of two distinct communities of news outlets. 
Furthermore, to better characterize inner group dynamics, we introduce a new technique which combines automatic topic extraction and sentiment analysis. 
We compare how the same topics are presented on posts and the related emotional response on comments finding significant differences in both echo chambers and that polarization influences the perception of topics. Our results provide important insights about the determinants of polarization and evolution of core narratives on online debating.
\end{abstract}


\end{frontmatter}

\section{Introduction}
The Arab Spring and Ukrainian revolution showed social media as a liberalizing technology and powerful vehicle of information, engagement, mobilization, able to encourage innovation and democracy. But social media have also changed the way we get informed and form our opinions. 

According to a recent report \cite{newman2015reuters}, approximately 63\% of users acquire their news from social media, and these news stories undergo the same popularity dynamics as other forms of online contents (such as selfies and cat photos). As a result of disintermediated access to information and of algorithms used in content promotion, communication has become increasingly personalized, both in the way messages are framed and how they are shared across social networks. 

Selective exposure and confirmation bias, indeed, have been shown to play a pivotal role in content consumption and information spreading \cite{del2016spreading}. Users tend to select information adhering (and reinforcing) their worldview and to ignore dissenting information\cite{quattrociocchi2016echo,bessi2015viral,zollo2015debunking,bessi2014science}. This pattern elicits the formation of polarized groups -- i.e., echo chamber -- where the interaction within like-minded people might even reinforce polarization \cite{zollo2015emotional,sunstein2002law}.

 Several studies pointed out the effects of social influence online \cite{centola2010,fowler2010cooperative,quattrociocchi2014opinion,salganik2006experimental}. Results reported in \cite{kramer2014} indicate that emotions expressed by others on Facebook influence our own emotions, providing experimental evidence of massive-scale contagion via social networks. 
Recent works \cite{bessi2014science,zollo2015debunking}, indeed, showed that attempts to debunk false information are largely ineffective. In particular, the discussion degenerates when the two polarized communities interact with one another. The increasing interest in online debates led researchers to investigate many of their aspects, from the characterization of conversation threads \cite{backstrom2013characterizing} to the detection of bursty topics on microblogging platforms \cite{diao2012finding}, to the disclosure of the mechanisms behind information diffusion for different kinds of contents \cite{romero2011differences}.

More recently, several doubts about social influence on the Internet have been raised during Brexit --the British referendum to leave the European Union-- campaign, where both sides, Leave and Remain, battled it out on social media. Indeed, a big effort has been dedicated to characterize the dynamics of the online Brexit debate, from applying filtering algorithms to study the shape of online data \cite{stolz2016topological}, through the investigation of the role of \textit{bots} on the direction of discussions \cite{howard2016bots}, to the study of the effects of the referendum result on financial markets \cite{bianchetti2016brexit}. 

In this paper we address the Brexit discussion on Facebook public pages referring to UK based official information sources listed in the European Media Monitor \cite{steinberger2013introduction}.  

Firstly, we characterize the structural properties of the discussion by observing the spontaneous emergence of two well-separated communities; indeed, connections among pages are the direct result of users' activity, and we do not perform any categorization of contents a priori. 
Then, we explore the dynamics behind discussion: looking at users polarization towards the two communities and at their attention patterns, we find a sharply bimodal distribution, showing that users are divided into two main distinct groups and confine their attention on specific pages.  

Finally, to better characterize inner group dynamics, we introduce a new technique which combines automatic topic extraction and sentiment analysis. We compare how the same topics are presented on posts and the related comments, finding significant differences in both echo chamber and that polarization on the perception of topics. We first measure the distance between how a certain concept is presented on the posts and then the emotional response of users to such controversial topics. Our new metrics could be of great interest to identify the most crucial topics in online debates. Indeed, it is highly likely that the greater the emotional distance between the same concept in two echo chambers, the greater the polarization of users involved in the discussion. Therefore, such a distance may be a key marker to locate controversial topics and to understand the evolution of the core narratives within distinct echo chambers. 

\section{Methods}
\subsection*{Ethics Statement.}
The data collection process was carried out using the Facebook
Graph API \cite{fb_graph_api}, which is publicly available. For the
analysis (according to the specification settings of the API) we only
used publicly available data (thus users with privacy restrictions are
not included in the dataset). The pages from which we downloaded data are
public Facebook entities and can be accessed by anyone. Users' content
contributing to such pages is also public unless users' privacy
settings specify otherwise, and in that case it is not available to us.
	
\subsection*{Data collection.}
The European Media Monitor (EMM) \cite{steinberger2013introduction} provides a list of all news sources which includes, for each of them, the related country and region. We limited our collection to all pages whose legal head office (at least one of them) is located in the United Kingdom. For each page, we downloaded all the posts from January 1st to July 15th, 2016, as well as all the related likes and comments. The exact breakdown of data is provided in Tab.~\ref{tab:dataset}, while the complete set of downloaded pages is reported in Tab.~\ref{tab:pages} in the \textit{Appendix}.

\begin{table}[ht]
	\centering	
	\begin{tabular}{|c |c |c| }\hline
		& \textbf{Total} &\textbf{Brexit}\\ \hline\hline
		\textit{Pages} & $81$ & $38$\\ 
		\textit{Posts} & $303,428$ & $5,039$\\
		\textit{Likes} &  $186,947,027$ &$2,504,956$ \\
		\textit{Comments} & $38,182,541$ & $469,397$ \\
		\textit{Likers} & $30,932,388$ & $1,365,821$\\
		\textit{Commenters} & $7,222,273$ & $259,078$\\\hline
	\end{tabular}\label{tab:dataset}
	\caption{\textbf{Dataset description:} from January 1st to July 15th, 2016.}	
\end{table}

\subsection*{Preliminaries and Definitions.}
In this section we provide a brief description of the main concepts and tools used in the analysis.

\subsubsection*{Bipartite Projection.}
A bipartite graph is a triple $\mathcal{G}=(A,B,E)$ where $A=\left\{ a_{i}\,|\,i=1\dots n_{A}\right\} $ and $B=\left\{ b_{j}\,|\,j=1\dots n_{B}\right\} $ are two disjoint sets of vertices, and $E\subseteq A\times B$ is the set of edges -- i.e. edges exist only between vertices of the two different sets $A$ and $B$. The bipartite graph $\mathcal{G}$ is described by the rectangular matrix $M$ defined as
\[
M_{ij}=\left\{ \begin{array}{cc}
1 & if\, an\, edge\, exists\, between\, a_{i}\, and\, b_{j}\\
0 & otherwise
\end{array}\right. .
\]

We consider the bipartite network $\mathcal{G}=(P,U,E)$ where $P$ is the set of Facebook pages concerned on Brexit topics (see Tab.~\ref{tab:pages} in the \textit{Appendix}) and $U$ is the set of users active on pages belonging to $P$. An interaction with a given information posted by a page $p\in P$ determines a link between a user $u \in U$ and the page $p$, hence $M_{p,u}=1$ indicates that user $u$ was active on page $p$. 
For our analysis we use the co-occurrence matrices $C^P=MM^{T}$ and $C^U=M^{T}M$ that count, respectively, the number of common neighbors between two vertices of $P$ or $U$. As an example, $C^P_{p,q}$ for $p\neq q$ counts the number of users that were active on both pages $p$ and $q$.  
$C^P$ can be interpreted as the weighted adjacency matrix of the co-occurrence graph $G^P$ with vertices on $P$. Each non-zero element $C^P_{p,q}$ corresponds to an edge $(p,q)$ among vertices $p$ and $q$ with weight $C_{pq}^P$.

\subsubsection*{Community Detection Algorithms.}
Community detection algorithms serve to identify groups of nodes in a network. Most of the strategies relies on the modularity which quantifies the division of a network in separated clusters, high modularity corresponds to  a dense connectivity between nodes in a community and sparse connections between modules. 
In this work we apply four different community detection algorithms: Fast Greedy (FG), that seeks for the maximum modularity score by considering all possible community structures in the network. It tries to optimize the modularity function in a greedy manner 
 \cite{clauset2004finding}. Walktrap (WT), that exploits the fact that a random walker tends to remain trapped in the denser part -- i.e., communities -- of a graph. Hence WT uses short random walks to merge separate communities \cite{pons2006computing}. 
Multilevel (ML), that  is based on a multi-level modularity optimization procedure 
\cite{blondel2008fast}. 
Label Propagation (LP) \cite{raghavan2007near}, that is a nearly linear time algorithm that gives unique labels to vertices that are then updated according to majority voting in the neighboring vertices. Dense group of nodes reach consensus on a common label quickly. 
To compare the various community structures we use standard methods that compute the similarity between different clustering methods by considering how nodes are assigned by each community detection algorithm \cite{rand1971objective,hubert1985comparing}.

\subsubsection*{Backbone Detection Algorithm.}
The disparity filter algorithm is a network reduction technique based on the local identification of the statistically relevant weight heterogeneities. This method is able to identify the backbone structure of a weighted network without destroying its multi-scale nature \cite{vespignani2009}. We make use of this algorithm to obtain the relevant connections that form our networks' backbones and produce clearer visualizations.

\section*{Results and Discussion}
As a preliminary step, we divide all UK based pages in two groups: \textit{Brexit pages}, that includes those pages engaged in the debate around the Brexit, and \textit{Non Brexit pages}.
Out of $81$ pages, $38$ posted at least one news story about the Brexit.
Hence, we characterize the users behavior on Brexit pages and their related posts.

\subsection{Communities and News Polarization}
Online social media proved to facilitate the aggregation of individuals in communities of interest, also known as \textit{echo chambers} \cite{quattrociocchi2016echo}, especially when restricting the interaction of users to conflicting information \cite{del2016spreading, zollo2015debunking}. In the case of Brexit pages we focus of the emerging communities without considering the shared contents, but rather by accounting for the connections created by users activities.

Therefore, we start by analyzing the community structure of the Brexit pages graph. We consider the bipartite projection of the pages-users graph $G_p$ where nodes are Brexit pages and two pages are connected if at least one user liked a post from each of them. The weight of a link is determined by the number of users in common between the two pages.

\begin{figure}[h]
 	\centering
 	\subfigure[]
 	{\includegraphics[width=0.7\textwidth]{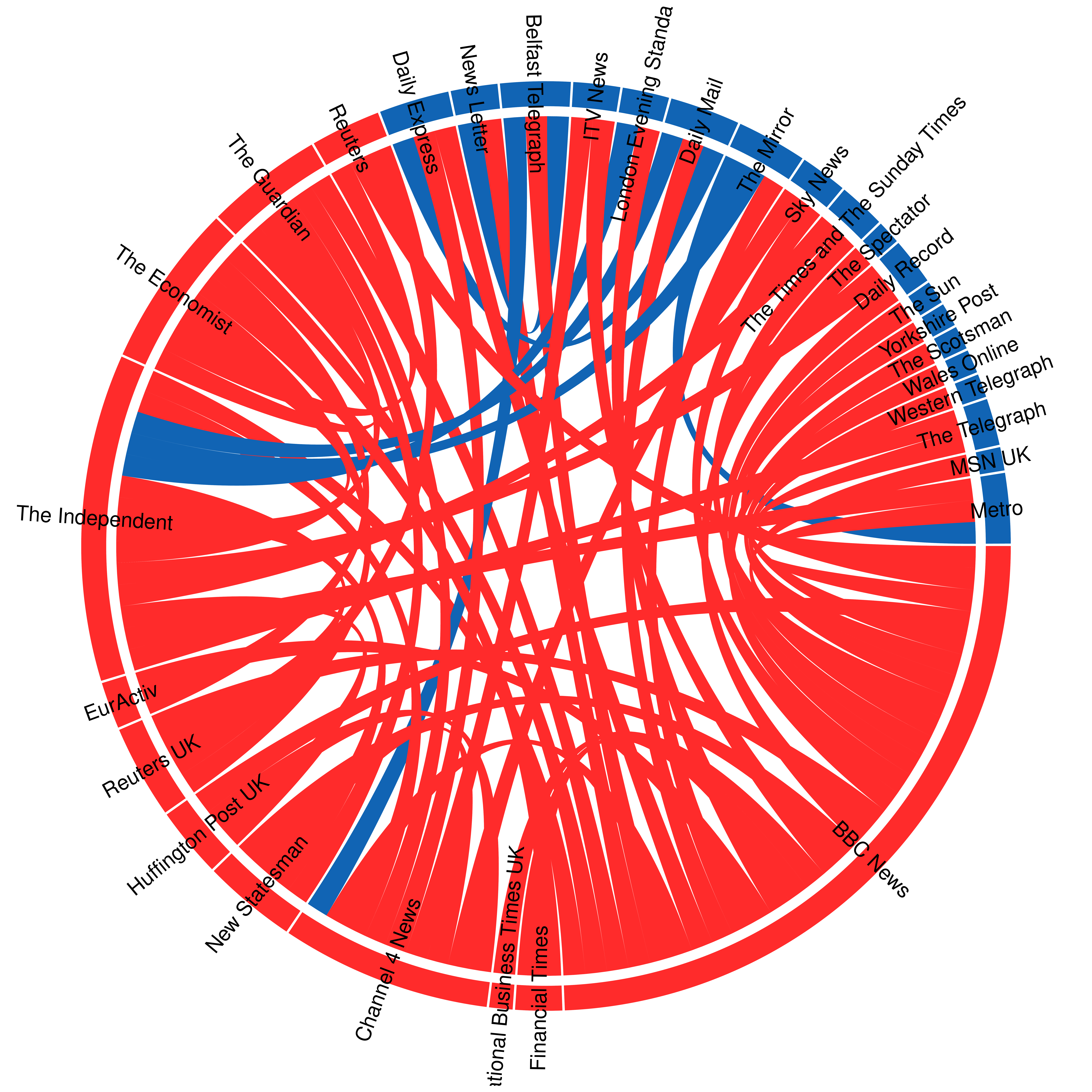}} 
 	\subfigure[]
 	{\includegraphics[width=0.2\textwidth]{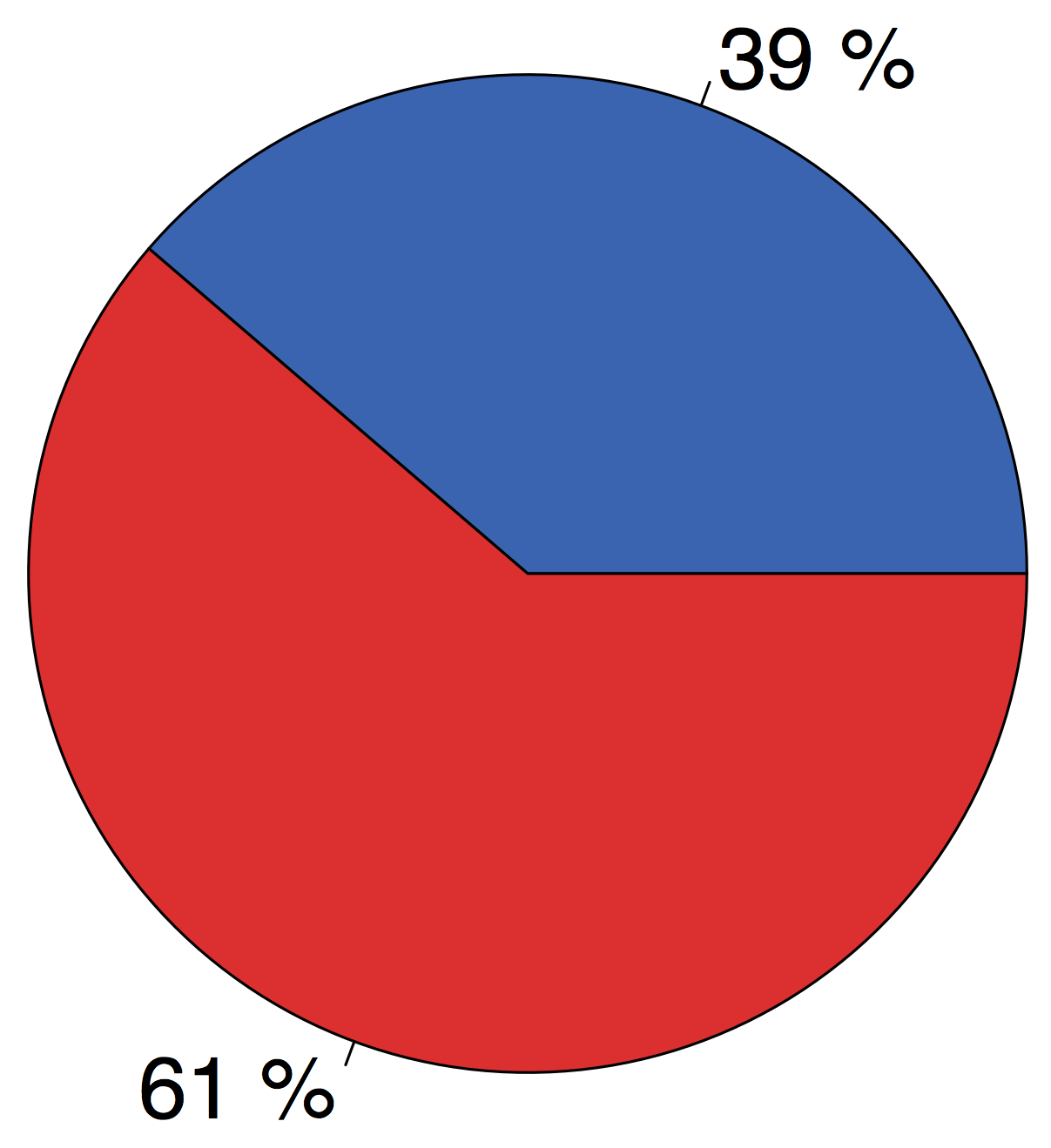}}
 	\caption{Backbone structure for the bipartite projection of the pages-users graph $G_p$ (a) and percentage of pages in the different communities (b). Colors indicate the membership of users in the different communities (\textit{blue} for C1, \textit{red} for C2) detected by the FG algorithm, while for the extraction of the backbone we considered the level of significance $\alpha = 0.03$.}\label{communities}
 \end{figure}

In Fig.~\ref{communities}(a) we show the backbone structure of $G_p$. Colors (resp., blue and red) represent the membership to one of the two communities (resp., C1 and C2) detected by the Fast Greedy (FG) algorithm (see \textit{Methods} section for further details). Fig.~\ref{communities}(b) reports the percentage of pages in both communities. A complete list of the pages and the relative membership is reported in Tab.~\ref{tab:pages} in the \textit{Appendix}. We compare the results of FG and two other community detection algorithms i.e., Walktrap (WT) and Multilevel (ML) (refer to \textit{Methods} section for further details) by means of the Rand method \cite{rand1971objective, hubert1985comparing}; we find a very high concordance between FG and ML (0.90), and lower ones between FG and WT (0.69), and between ML and WT (0.63).
Our analysis underlines the spontaneous emergence of two separate communities active on Brexit pages, where the connections among pages are a simple result of the interaction of users on them.

\begin{figure}
	\includegraphics[scale = .5]{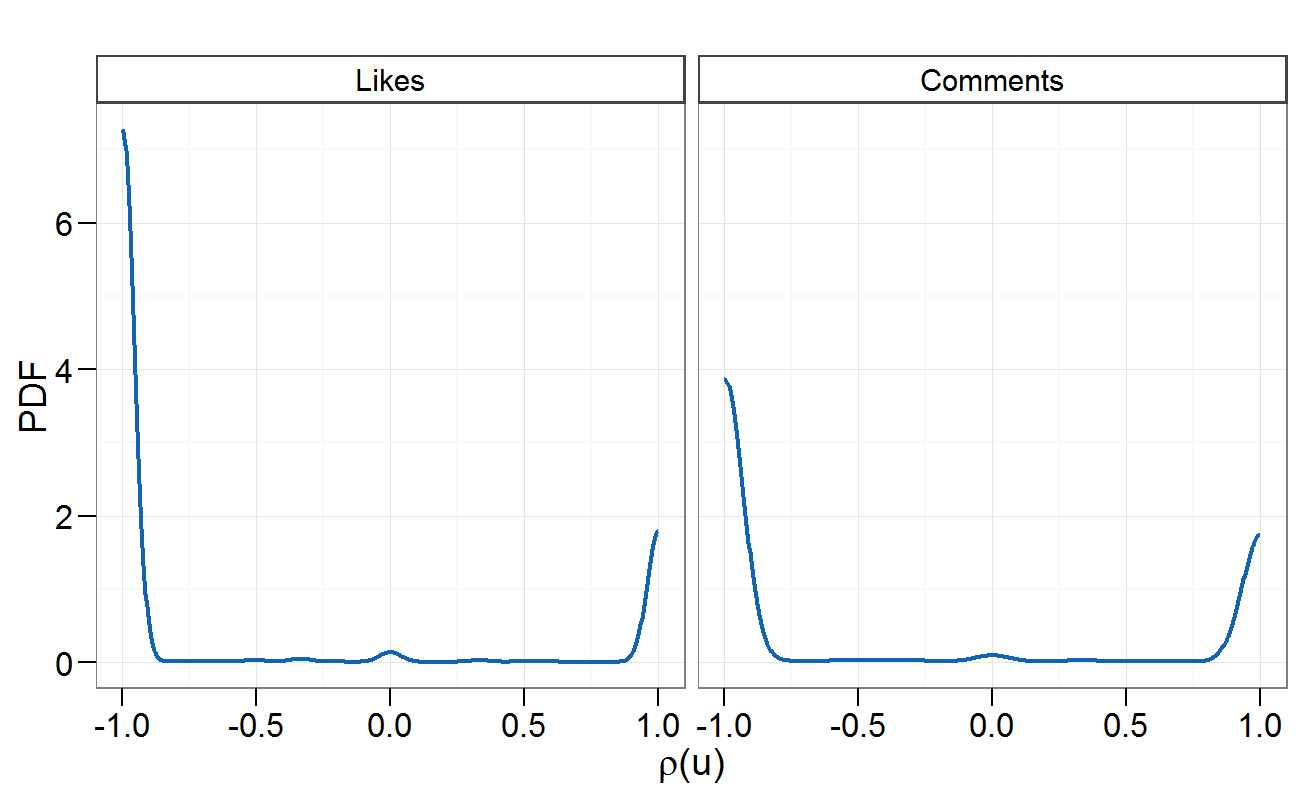}
	\caption{Probability density function (PDF) for the
		users polarization $\varrho(u)$ by likes \textit{(left)} and by comments \textit{(right)}. $\varrho(u) = 1$ (resp. $\varrho(u) = -1$) indicates that users $u$ is polarized towards C2 (resp., C1).  }\label{polarization}
\end{figure}

Taking into account the positive meaning of the like as a feedback to a post, we characterize how contents from the two communities detected in $G_p$ are consumed by Facebook users. 
We define the users polarization by likes (reps., comments) as
$$
\varrho(u) = (y - x)/(y + x),
$$
where $y$ is the number of likes (resp., comments) that user $u$ left on posts of C2 and $x$ the number of likes (resp., comments) left on posts of C1. Thus, a user $u$ is said to be polarized towards C2 (resp., C1) if $\varrho(u) = 1$ (resp., $ -1$). In Fig. \ref{polarization} we report the Probability Density Function (PDF) of users polarization by likes (left panel) and comments (right panel). 
We find that $\varrho(u)$ is sharply bimodal in both cases, denoting that the majority of users may be divided into two main groups referring to the two communities of Fig.~\ref{communities}(a). Tab.~\ref{tab:polarized} shows the number of polarized users towards both communities by likes and comments.

\begin{table}[ht]
	\centering	
	\begin{tabular}{|l|c|c|}\hline
		& \textbf{C1}&\textbf{C2}\\ \hline\hline
		\textit{Likes} & $1,037,969$ & $255,930$\\
		\textit{Comments} & $168,680$ & $75,851$\\ \hline
		\end{tabular}
	\caption{Number of polarized users towards both communities by likes and comments.}\label{tab:polarized}	
\end{table}	
	
\begin{figure}
	\centering
	\subfigure[]
	{\includegraphics[width=0.85\textwidth]{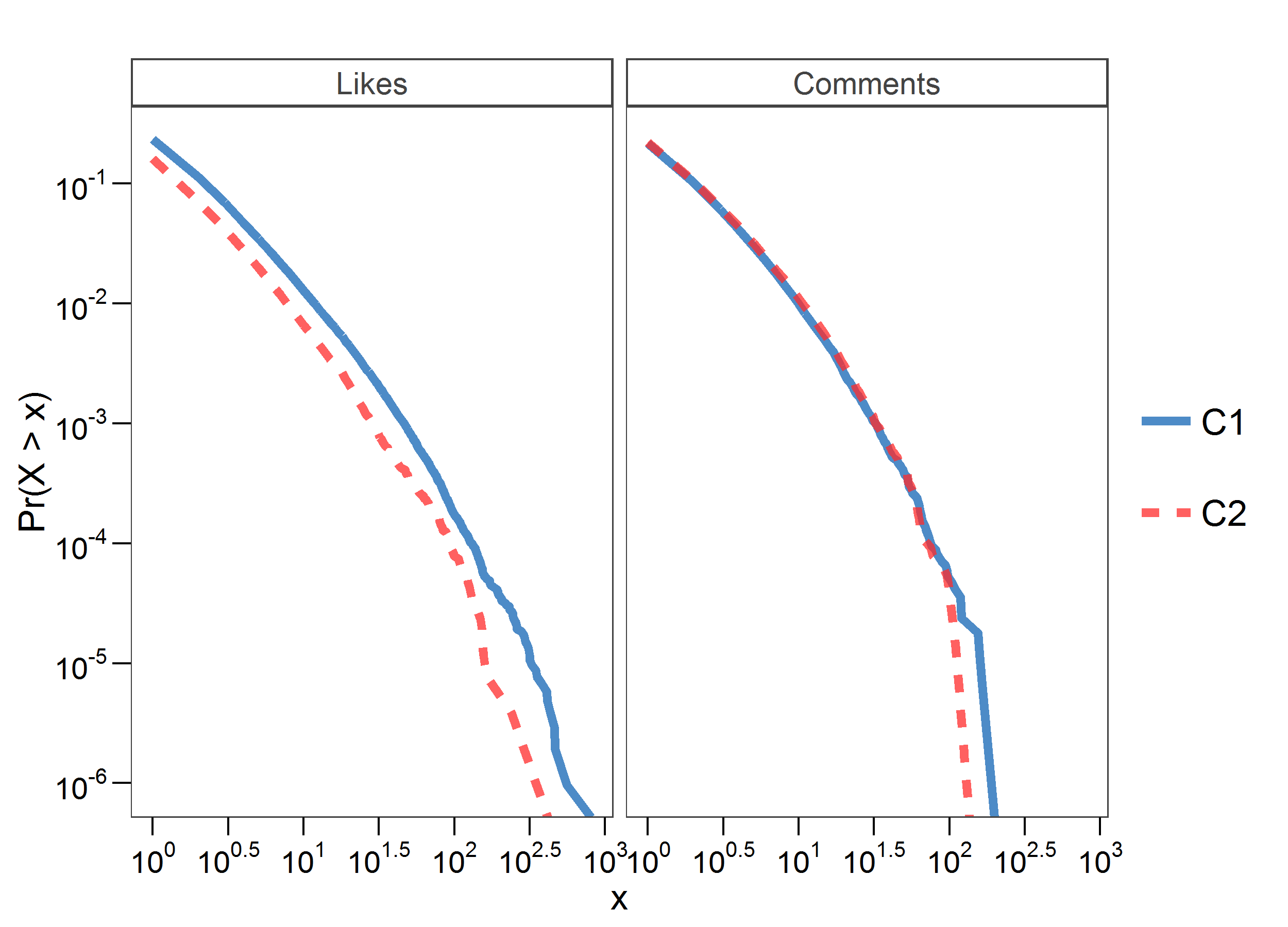}}
	
	\centering
	\subfigure[]
	{\includegraphics[width=0.65\textwidth]{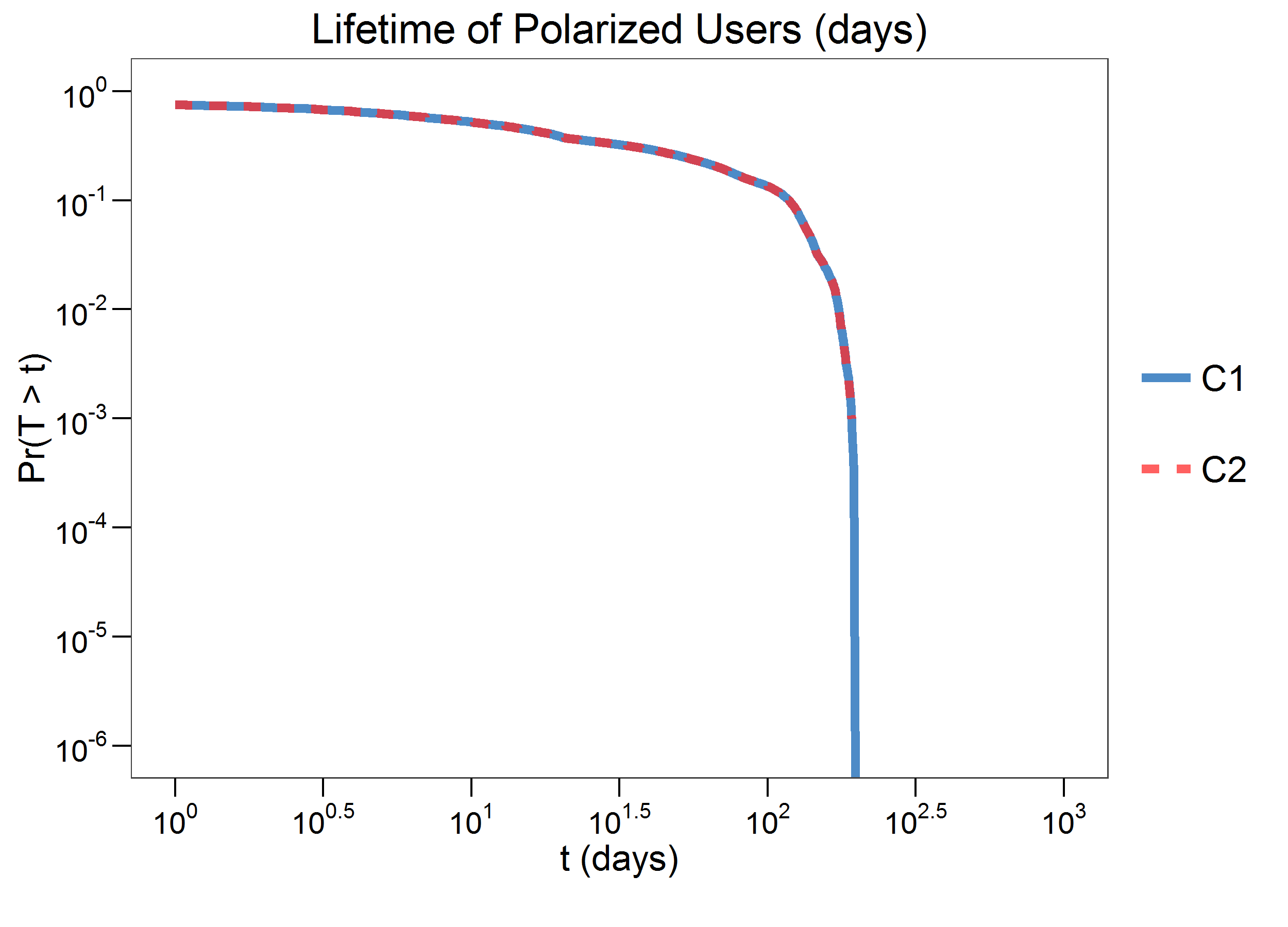}} 
	\caption{(a) CCDF of likes \textit{(left)} and comments \textit{(right)} made by users polarized towards C1 \textit{(solid blue)} and towards C2 \textit{(dashed red)}. (b) CCDF of the lifetime of users polarized towards either C1 \textit{(solid blue)} or 2 \textit{(dashed red)}. The lifetime is computed as the temporal distance, in terms of days, between the first and last comment made by any given user. }\label{polarized_users}
\end{figure}

Thus, we have shown that users form two well segregated communities. We now want to compare their activities on posts from Brexit pages. In Fig.~\ref{polarized_users}(a) we report the Complementary Cumulative Distribution Function (CCDF) of likes (left) and comments (right) made by users polarized towards both communities, while in Fig.~\ref{polarized_users}(b) we report the CCDF of the lifetime of polarized users. The lifetime is defined as the temporal distance, in terms of days, between the first and last comment made by any given user.
We fitted the distributions in Fig.~\ref{polarized_users}(a) with different models (the exponential, the power law, and the log-normal) by means of NLS estimation, goodness of fit tests are based on the maximization of the log-likelihood. Then, we pairwise compare the distributions of the number of likes and comments by users polarized towards either community by means of the \textit{Kolmogorov-Smirnov} (KS) test. Results for the best fit and KS test are reported in Tab.~\ref{tab:pol_users}. We may notice that the distributions are all best fitted by the exponential model, with the exception of that of the number of likes by users from C2, that is best fitted by the power-law. Also, we fail to reject the null hypothesis of equivalence of the two distributions in the case of the number of comments by users from either community.

\begin{table}[ht]
	\centering
	\begin{tabular}{|c |c | c|}\hline
		\textbf{Distribution} & \textbf{Best Fit} &\textbf{Estimated Parameters}\\ \hline \hline
		\textit{$\#$ of likes (C1)} & exponential&$ \hat{a} = 1.49, \, \hat{b} = 0.00003$ \\ 
		\textit{$\#$ of likes (C2)} & power-law & $ \hat{a} = 1.18, \, \hat{b}= -0.03 $\\
		\textit{$\#$ of comments (C1)} & exponential & $\hat{a} = 1.55, \, \hat{b} = 0.00005$\\
		\textit{$\#$ of comments (C2)} &  exponential &$\hat{a} =1.58,\, \hat{b} =0.0001$\\ \hline
		\multicolumn{3}{c}{ }\\ \hline
		\multicolumn{3}{|c|}{\textbf{KS test}}\\ \hline\hline
		\textbf{Compared Distributions} & \textbf{D (C) }&\textbf{ p-value}\\ \hline \hline	
		\textit{$\#$ of likes (C1/C2)} & $ 0.075\,(0.004) $ &  $2\times10^-16$ \\
		\textit{$\#$ of comments (C1/C2)} & $\mathbf{ 0.004}\, (0.007) $& $ 0.349$ \\ \hline
		\end{tabular}
		\caption{\textbf{Fit of distributions} from Fig. \ref{polarized_users} and results of Kolmogorov-Smirnov tests.}		\label{tab:pol_users}
\end{table}

Interestingly, although users tend to focus on contents coming just from one of the two communities, the distributions of their attention patterns are very similar, and even equal in the case of comments.

\subsection{Emotional Dynamics Inside and Between Communities}
Our analysis provides evidence of the existence of two well segregated echo chambers: users tend to focus on one narrative and to ignore the other. Such a pattern might be driven by the way in which contents are debated on pages, i.e., is such a way that matches their own users' preferences. To shade light on this aspect, we want to measure the distance among the sentiment of the users w.r.t. the same topic. Thus, we analyze how the subject of a post is presented to the users. To perform the analysis we make use of IBM Watson\texttrademark~AlchemyLanguage service API \cite{alchemy}, that allows us to extract semantic meta-data from posts content. Such a procedure applies machine learning and natural language processing techniques aimed to analyze text by automatically extracting relevant entities, their semantic relationship as well as the emotional sentiment they express \cite{gangemi2013comparison}. 
In particular, we extract the sentiment and the main concepts presented by each post of the dataset, whether it has a textual description or a link to an external document. The AlchemyAPI tools make use of the language patterns surrounding the input text looking for signals that denote the sentiment and exploring information based on the concepts behind such an input. Thus, a concept is a high-level conceptual association identified in the content provided as input to the service. Input content is auto-tagged against a concept graph, which formally represents the relationships between the concepts contained in the data on which it is based. 

\begin{figure}[h!]
\centering
	\includegraphics[scale = .23]{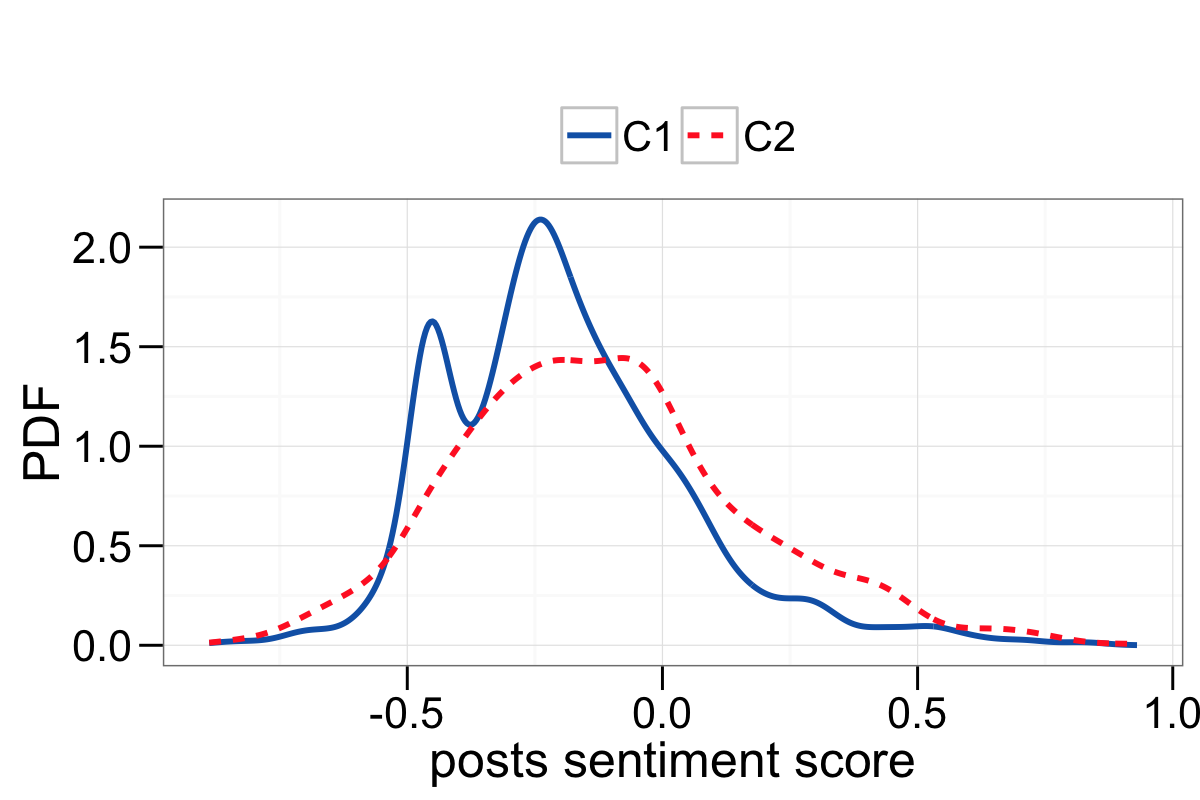}
	\caption{Probability Density Function (PDF) of posts sentiment score on C1 \textit{(solid blue)} and C2 \textit{(dashed red)}. The sentiment score is defined in the range $[-1,1]$, where $-1$ is negative, $0$ is neutral, and $1$ is positive}\label{sentiment_post}
\end{figure}	

Fig.~\ref{sentiment_post} shows the sentiment distribution of posts on both communities. The sentiment score is defined in the range $[-1,1]$, where $-1$ is negative, $0$ is neutral, and $1$ is positive. We may observe a  negative overall pattern for both categories, although clearly more pronounced for posts of C1. Notice that we consider how subjects are presented in a post; here we do not take into account the sentiment that the post may elicit in the reader, or the sentiment of users involved in the discussion.

\subsubsection*{Controversial Concepts: Emotional Distance and Users' Response}
\begin{figure}
	\centering
	\subfigure[]
	{\includegraphics[width=1\textwidth]{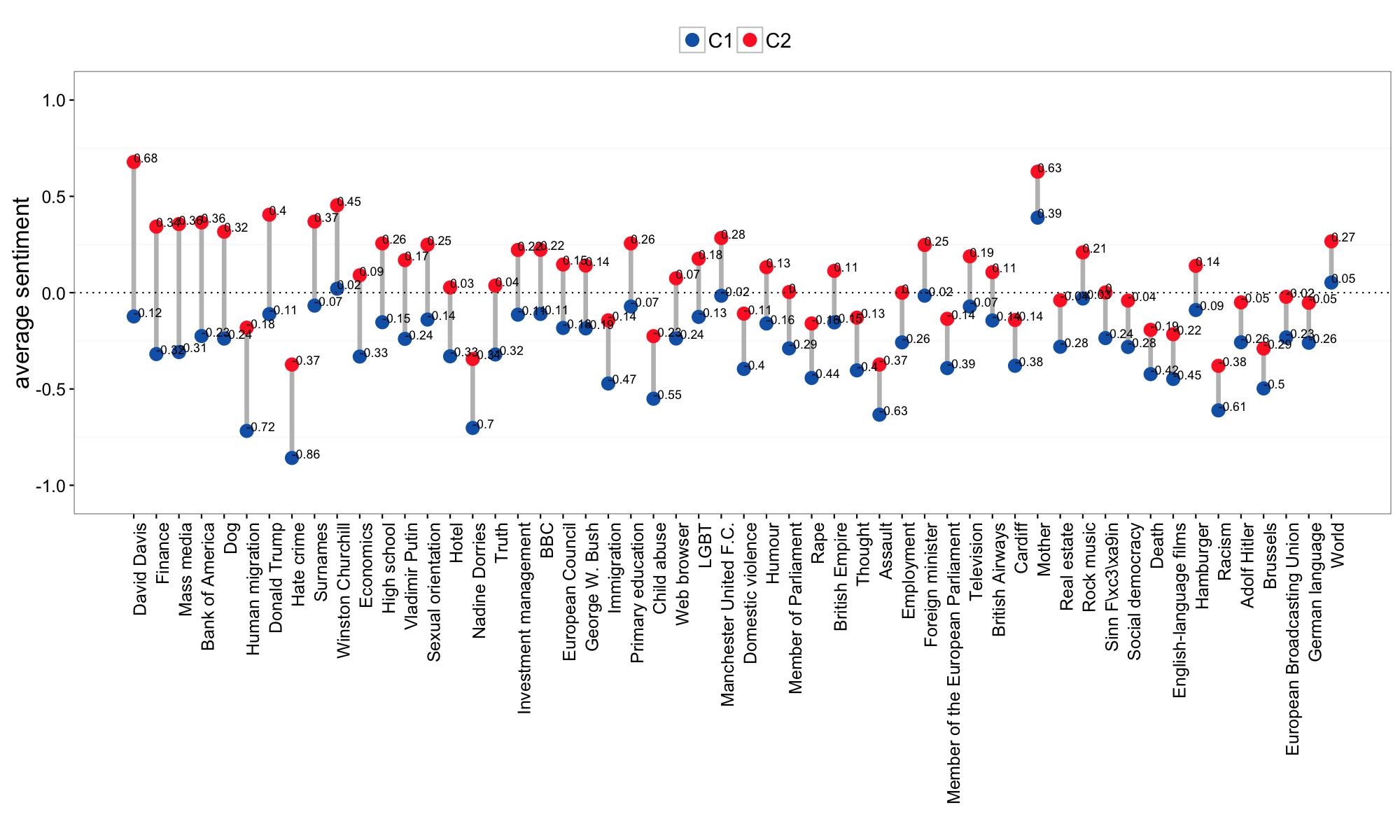}} 
	\subfigure[]
	{\includegraphics[width=1\textwidth]{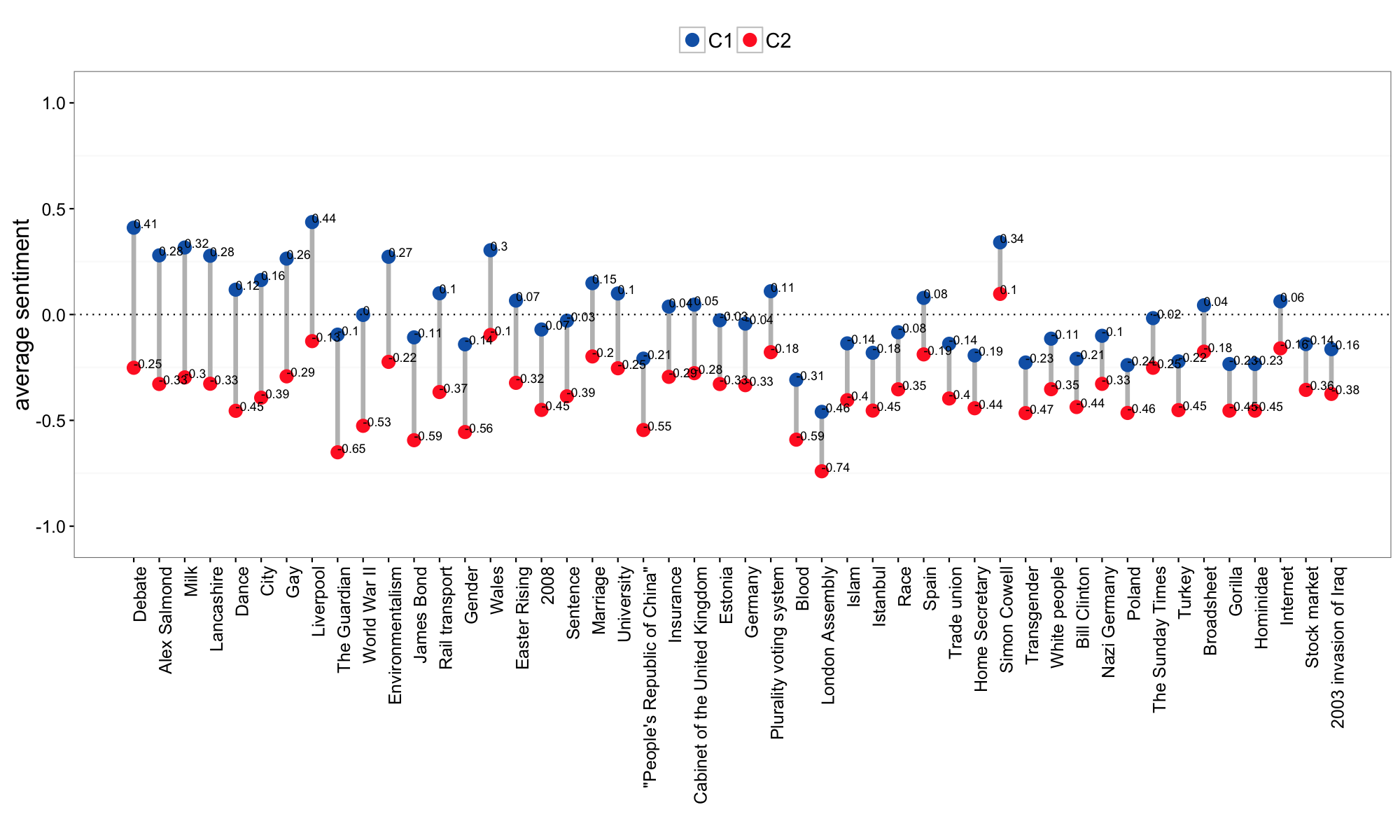}}
	\caption{\textbf{Emotional Distance Between Communities.} Emotional distance -- i.e., the distance between the average sentiment of a concept on both communities -- for each concept debated in both communities. Panel (a) includes the $52$ concepts that are presented in a more negative way in community C1 (blue dots) w.r.t. C2 (red dots), while panel (b) includes the $48$ concepts that are presented in a more negative way in community C2 (red dots) w.r.t. C1 (blue dots). Concepts are shown in a descending order by the largest to the smallest emotional distance.}
	\label{concept_diff}
\end{figure}

We now want to understand how users of the two echo chambers perceive the issues debated on their pages. Thus, we focus on the top-100 concepts of each echo chamber: 102 such concepts are shared by both communities, for a total of $1,520$ posts ($1,258 \in C1$, $262 \in C2$) and $115,958$ comments ($95,357 \in C1$, $20,601 \in C2$). For each concept we compute its average sentiment -- i.e., the mean of the sentiment of all the posts where it appears. The emotional distance between two concepts is defined as the difference between the average sentiment of the concept in $C2$ and that in $C1$. Since we are interested in identifying the most controversial concepts, we consider only those concepts for which the emotional distance (in absolute value) between the two communities is greater than $0.2$. Fig.~\ref{concept_diff} shows, for each concept, the emotional distance between the two echo chambers. More specifically, the top panel (a) of Fig.~\ref{concept_diff} includes the $52$ concepts that are presented in a more negative way in community C1 w.r.t. C2, while the bottom panel (b) includes the $48$ concepts that are presented in a more negative way in community C2 w.r.t. C1. In both panels concepts are shown in descending order by the largest to the smallest emotional distance. Thus, concepts on the left are those discussed with the greatest difference in sentiment, while those on the right are discussed in a much more similar way by both echo chambers.

\begin{figure}
	\centering
	\subfigure[]
	{\includegraphics[width=\textwidth]{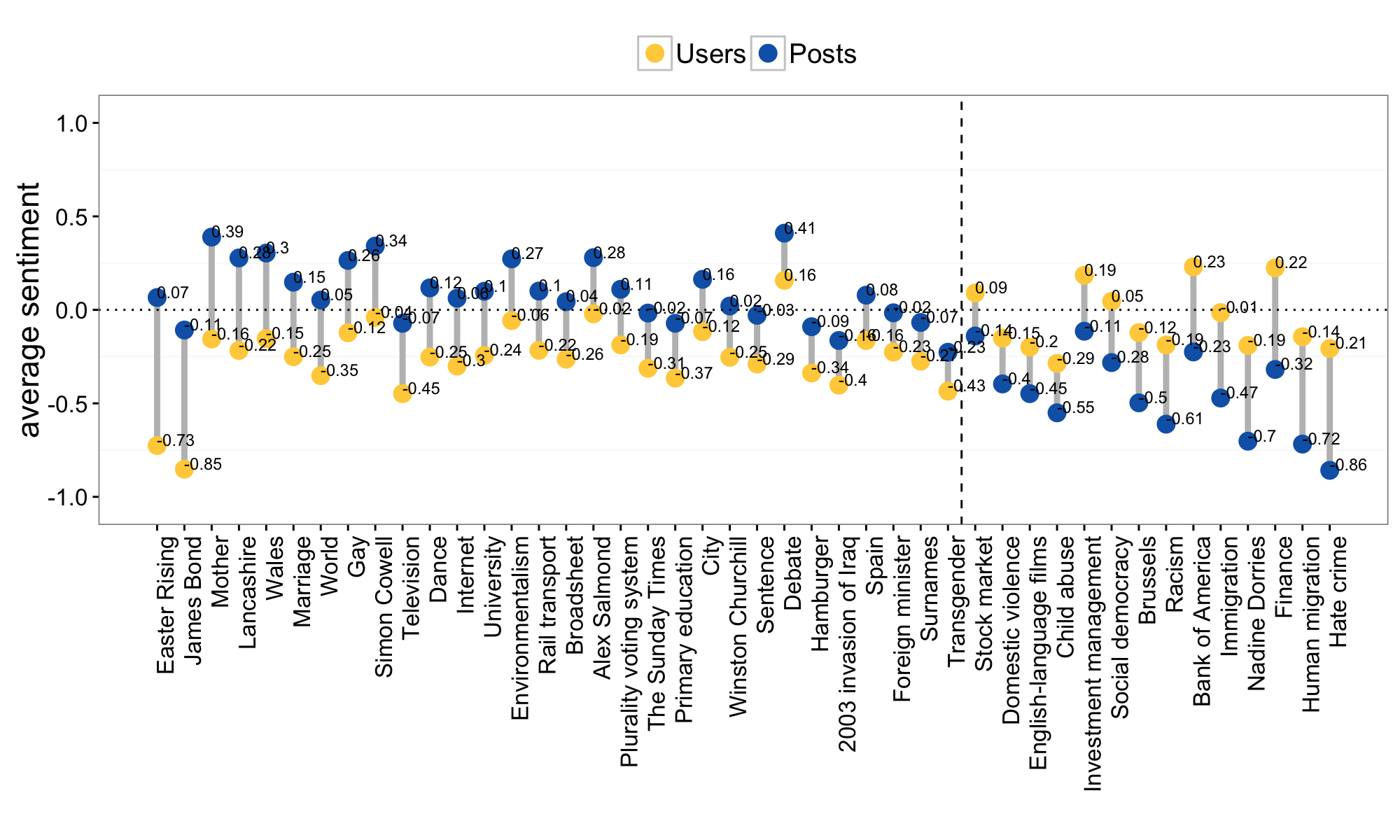}} 
	\subfigure[]
	{\includegraphics[width=\textwidth]{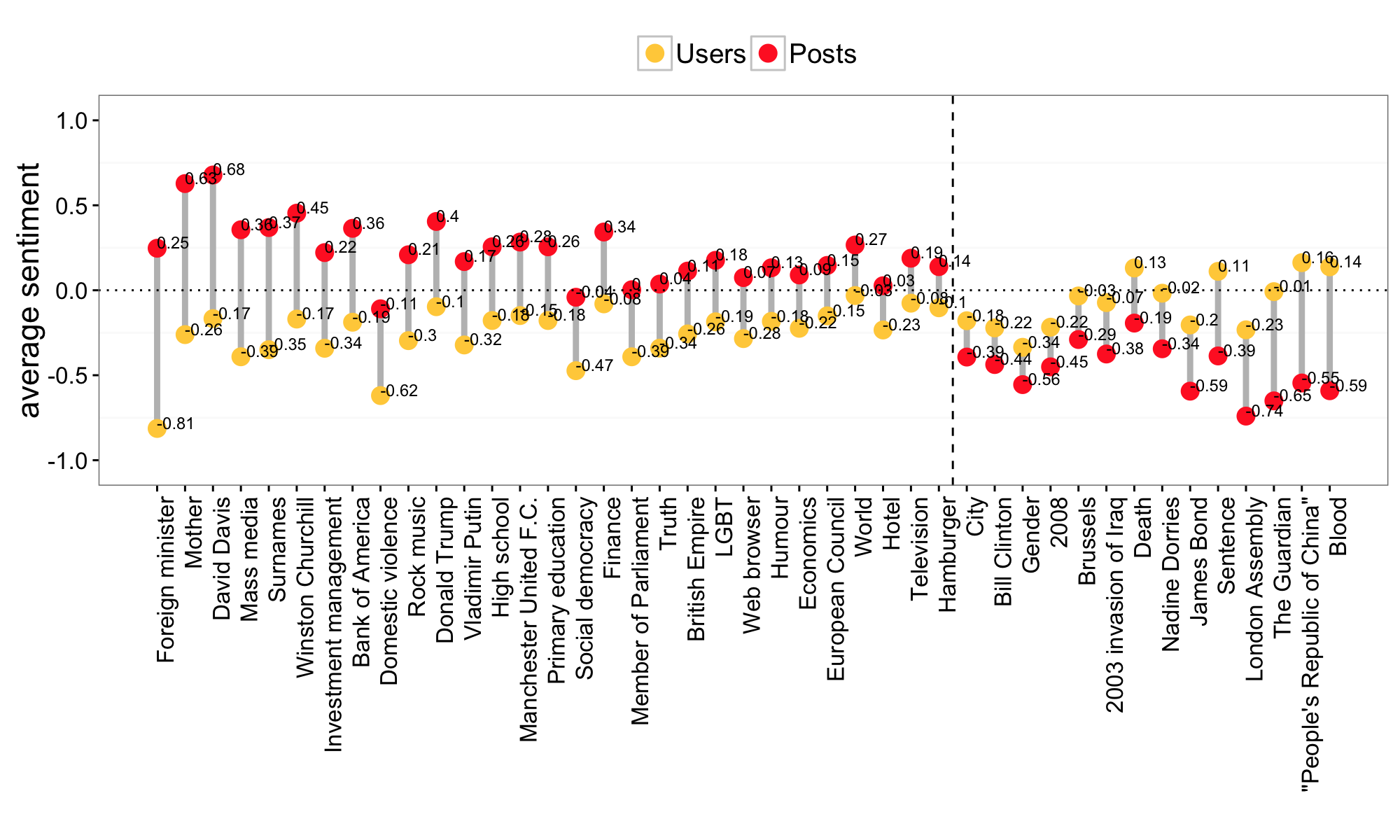}} 
	\caption{\textbf{Emotional Response to Controversial Concepts.} Panel (a) shows the emotional response of users (yellow dots) to posts of C1 (blue dots) debating one of the listed controversial concepts, while panel (b) shows the emotional response of users (yellow dots) to posts of C2 (red dots). Only concepts for which the emotional distance between the two communities is greater than $0.2$ are reported. The vertical dashed lines denote a change in users' response. }
	\label{post_user}
\end{figure}

So far we have analyzed how subjects are debated in the posts of Brexit pages. What about the emotional response of users to such posts? To this aim we take all the comments ($115,958$) of posts including one of the top-100 concepts and compute their sentiment score through AlchemyAPI. Thus, to each comment is associated a sentiment score in $[-1,1]$, where $-1$ is negative, $0$ is neutral, and $1$ is positive. For each post (resp., user) we compute the average sentiment of its (resp., her) comments -- i.e., the mean of the sentiment of all comments on the post (resp., made by the user). Then, for each concept, we consider the emotional distance between the average sentiment of the post and that of its users. Fig.~\ref{post_user} shows the emotional response of users to posts of C1 (a) and C2 (b) debating one of the listed controversial topics. Only concepts for which the emotional distance (in absolute value) between the two communities is greater than $0.2$ have been taken into account. In both panels a vertical dashed line denotes a change in users' response: concepts on the left are those for which users' response is more negative than the sentiment expressed in the post, and vice versa for those on the right. We may notice that users tend to react negatively to the content of the posts, independently of their reference community. 

\section*{Conclusions}
We address the online discussion around Brexit on Facebook by means of a quantitative analysis on a sample of 5K posts from 38 pages linked to official UK news sources.
We observe the spontaneous emergence of two separate communities, where the connections among pages are the direct result of users' activity and no reference to the shared contents is implied. We further explore the dynamics of the discussion by looking at the polarization of users from the two communities and their attention patterns. We find a sharply bimodal distribution for the polarization of users. Users segregation might be driven by the match between their personal preferences and the way in which contents are presented.
We identify how concepts get received and shape the narrative inside the echo chamber by measuring both the distance between the sentiment of users w.r.t. the same topic and that of users w.r.t. the ``presentation'' of the topic. Firstly, we characterize the structural properties of the discussion by observing the spontaneous emergence of two well-separated communities; indeed, connections among pages are the direct result of users' activity, and we do not perform any categorization of contents a priori. 
Then, we explore the dynamics behind discussion: looking at users polarization towards the two communities and at their attention patterns, we find a sharply bimodal distribution, showing that users are divided into two main distinct groups and confine their attention on specific pages.  Finally, to better characterize inner group dynamics, we introduce a new technique which combines automatic topic extraction and sentiment analysis. We compare how the same topics are presented on posts and the related comments, finding significant differences in both echo chambers and that polarization reflects on the perception of topics. We first measure the distance between how a certain concept is presented on the posts by both echo chambers and then we measure the emotional response of users to such controversial topics. Our new measures could be of great interest to identify the most crucial topics in online debates. Indeed, it is highly likely that the greater the emotional distance between the same concept in two echo chambers, the greater users' polarization. 
Our results provide important insights for identifying the determinants of polarization and evolution of the core narratives behind online debating.

\section*{References}
\bibliography{mybib}

\section*{Appendix}
In this section we provide the list of all Facebook pages of news sources whose legal head office (at least one of them) is located in the United Kingdom. Pages engaged in the debate around Brexit are denoted by Y (N, otherwise), followed by the community to which they belong (C1 or C2).

\begin{small}
\begin{longtable}{|l|l|r|c|} \hline
\textbf{ID} & \textbf{Page Name} & \textbf{Facebook Code} & \textbf{Brexit}\\ \hline\hline
1 & BBC News  & 228735667216  & Y(C1)\\
2 & Channel 4 News  & 6622931938  & Y(C1)\\
3 & Euractiv (English)  & 15299247059  & Y(C1)\\
4 & Financial Times  & 8860325749  & Y(C1)\\
5 & Huffingtonpost UK  & 143753582359049  & Y(C1)\\
6 & International Business Times UK  & 224377357631653  & Y(C1)\\
7 & New Economics Foundation  & 110275553302  & Y(C1)\\
8 & New Statesman  & 100959719644  & Y(C1)\\
9 & Open Europe Today  & 321253057971308  & Y(C1)\\
10 & Reuters  & 114050161948682  & Y(C1)\\
11 & Reuters UK  &208314602512037 & Y(C1)\\
12 & The Economist  & 6013004059  & Y(C1)\\
13 & The Guardian UK  & 10513336322  & Y(C1)\\
14 & The Independent  & 13312631635  & Y(C1)\\
15 & The Register  & 206419956048907  & Y(C1)\\
16 & WN.com  & 229101503845879  & Y(C1)\\
17 & Belfast Telegraph  & 237692023818  & Y(C2)\\
18 & Daily Express  & 129617873765147  & Y(C2)\\
19 & Daily Mail   & 164305410295882  & Y(C2)\\
20 & Daily Record  & 187523381277554  & Y(C2)\\
21 & East Anglian Daily Times  & 6478299951  & Y(C2)\\
22 & ITV News  & 148007467671  & Y(C2)\\
23 & London Evening Standard  & 165348596842143  & Y(C2)\\
24 & Metro (UK) & 117118184990145  & Y(C2)\\
25 & MSN UK  & 358837740527  & Y(C2)\\
26 & News Letter  & 117370764948881  & Y(C2)\\
27 & Nottingham Post  & 309833935716287  & Y(C2)\\
28 & Sky News  &164665060214766 & Y(C2)\\
29 & The Mirror  & 6149699161  & Y(C2)\\
30 & The Scotsman  &293226174987 & Y(C2)\\
31 & The Spectator  & 111263798903232  & Y(C2)\\
32 & The Sun  & 161385360554578  & Y(C2)\\
33 & The Times and Sunday Times  & 147384458624178  & Y(C2)\\
34 & Wales Online  & 21226447182  & Y(C2)\\
35 & Wandsworth Guardian  & 113349742029506  & Y(C2)\\
36 & Western Telegraph  & 180521675319022  & Y(C2)\\
37 & Yorkshire Post  & 316795048375439  & Y(C2)\\
38 & Airforce Technology Website  & 376588539031515  & N\\
39 & Azo Mining  & 195005930530874  & N\\
40 & BBC Radio  & 1470145583204820  & N\\
41 & Cafebabel (English)  & 357343795001  & N\\
42 & City A.M.  & 213682385348579  & N\\
43 & Dunmow Broadcast  & 181182540669  & N\\
44 & EU business  & 215108901846669  & N\\
45 & Euromoney  & 192279900885723  & N\\
46 & European Railway Review  & 404359882930504  & N\\
47 & Expatica  & 206982432584  & N\\
48 & Farming Life  & 243070359106664  & N\\
49 & FCO - Foreign and Commonwealth Office  & 408582579294175  & N\\
50 & Harborough Mail  & 219817851378553  & N\\
51 & Herald Scotland  & 271154343382  & N\\
52 & Inmarsat  & 317156988374684  & N\\
53 & Lydian International  & 186900121339682  & N\\
54 & Mining Technology  & 326019370778750  & N\\
55 & Mondo Visione  & 169767016460715  & N\\
56 & MoneyWeek  & 110326662354766  & N\\
57 & Monsters and Critics  & 193326863118  & N\\
58 & New Civil Engineer  & 166793706822441  & N\\
59 & OneWorld.net: Palestine  & 106968052697581  & N\\
60 & Oxford Analytica  & 160525917321265  & N\\
61 & Pan European Networks  & 230201663697109  & N\\
62 & Publish What You Pay  & 176624229034172  & N\\
63 & Railway Magazine  & 135345903226042  & N\\
64 & Routes News  & 126251777434574  & N\\
65 & Seatrade Global  &470795739645931 & N\\
66 & Survival International  & 19668531552  & N\\
67 & Tax-News  & 375456009146619  & N\\
68 & The Argus  & 57197526698  & N\\
69 & The Courier  & 325681791214  & N\\
70 & The International Institute for Strategic Studies  & 29840385993  & N\\
71 & The Scottish Government  & 200786289976224  & N\\
72 & The Telegraph  & 143666524748  & N\\
73 & The Visitor  & 68554461041  & N\\
74 & This is Africa  & 779213412106756  & N\\
75 & This is Derbyshire  & 142370589115824  & N\\
76 & This is Staffordshire & 11878899813  & N\\
77 & Thomson Reuters Foundation & 31301735406  & N\\
78 & World Fishing - The Magazine for Fishing  & 552321618120006  & N\\
79 & Cyprus Expat News  & 357342727764507  & N\\
80 & African Business Magazine  &114117578656259 & N\\
81 & African Review  &507239115959583 & N\\  \hline
\caption{\textbf{UK Facebook News Sources and Brexit Community Membership}: List of all Facebook pages of news sources whose legal head office (at least one of them) is located in the United Kingdom. Pages engaged in the debate around Brexit are denoted by \textit{Y}, followed by the community to which they belong.}
\label{tab:pages}
\end{longtable}
\end{small}

\end{document}